\documentclass[doublecol]{epl2_format} 

\usepackage{graphicx}
\usepackage{amsmath}
\usepackage{amssymb}
\usepackage{epstopdf}

\title{Lateral diffusion of receptor-ligand bonds in membrane adhesion zones: Effect of thermal membrane roughness} 
\shorttitle{\small Lateral diffusion of receptor-ligand bonds in membrane adhesion zones}
\author{H.\ Krobath$^{1,2}$ \and G.\ J.\ Sch\"utz$^{2}$ \and R.\ Lipowsky$^{1}$ \and T.\ R.\ Weikl$^{1}$}
\shortauthor{H.\ Krobath, G.\ J.\ Sch\"utz, R.\ Lipowsky, and T.\ R.\ Weikl}
\institute{
 \inst{1}Max Planck Institute of Colloids and Interfaces,  14424 Potsdam, Germany \\
 \inst{2}University Linz, Altenbergerstr.\ 69, 4040 Linz, Austria
 }

\abstract{
The adhesion of cells is mediated by membrane receptors that bind to complementary ligands in apposing cell membranes. It is generally assumed that the lateral diffusion of mobile receptor-ligand bonds in membrane-membrane adhesion zones is slower than the diffusion of unbound receptors and ligands. We find that this slowing down is not only caused by the larger size of the bound receptor-ligand complexes, but also by thermal fluctuations of the membrane shape. We model two adhering membranes as elastic sheets pinned together by receptor-ligand bonds and study the diffusion of the bonds using Monte Carlo simulations. In our model, the fluctuations {\em reduce} the bond diffusion constant in planar membranes by a factor close to 2 in the biologically relevant regime of small bond concentrations. 
}

\begin{document}

\maketitle

\section{Introduction}

Advances in optical microscopy have made it possible to observe the diffusion of single lipid and protein molecules in biological membranes with a positional accuracy down to about 40 nm at ms time resolution \cite{Sako00,Schuetz00}. Deviations from free Brownian motion were frequently found \cite{Schuetz00,Lommerse05} and interpreted as the consequence of membrane microheterogeneity. Moreover, cis-interactions of mobile membrane proteins with cytoskeletal-anchored proteins or aggregates were identified via transient immobilization \cite{Douglass05}. Trans-interactions between single cellular cadherin molecules and their soluble counterpart have been studied \cite{Baumgartner03}, paving the way for a characterization of { membrane-membrane} adhesion zones at the level of individual adhesion molecules. 

Current single-molecule experiments aim at detecting the reduced diffusion constants of bound receptors and ligands as important parameters that report on the time course of interaction in { membrane adhesion} zones. It is generally expected that the diffusion of bound receptor-ligand complexes is slower than the diffusion of the constituent unbound receptor and ligand molecules, simply because the complexes are larger and therefore also exposed to a larger viscous drag. We focus here on another and, to our knowledge, so far neglected aspect of the lateral diffusion of receptor-ligand bonds: The coupling of the diffusion to thermal membrane shape fluctuations. We show that shape fluctuations on a nanometer scale significantly decrease the diffusion constant of the bonds. On this length scale, the fluctuations are governed by the bending energy of the membranes.

We consider a statistical-mechanical model in which two apposing membranes are discretized into quadratic patches. Each of the patches in the first membrane can be occupied by a single receptor molecule, which can bind a ligand molecule in the apposing patch of the second membrane. We consider here irreversibly bound receptor-ligand bonds that impose a local membrane separation close to the bond length $l_o$. The lateral diffusion of the bonds is modeled as a hopping process to neighboring patches and implies that the local separation at these patches is changed to $l_o$. Hopping steps therefore can lead to an increase of the membrane bending energy, which slows down diffusion.  

We show that the decrease of the diffusion constant depends on the thermal membrane roughness, which is a measure for the strength of fluctuations. The roughness, in turn, depends on the concentration of the receptor-ligand bonds: it decreases with increasing bond concentration. For small bond concentrations, the membrane roughness at neighboring membrane patches of receptor-ligand bonds tends towards a finite value. { In our statistical-mechanical model}, the fluctuations reduce the diffusion constant by a factor of about 1.8 in the regime of small bond concentrations, which is the biologically relevant regime for typical concentrations of receptor-ligand bonds in { membrane-membrane} adhesion zones.

\section{Model}

We consider two on average parallel membranes that are interconnected by receptor-ligand bonds. In our model, the membranes are discretized into quadratic patches of size $a\times a$. The linear patch size $a$ corresponds to the smallest bending deformations, which are around 5 nm, according to simulations with molecular membrane models \cite{Goetz99}. Each patch $i$ in the first membrane directly apposes a patch in the second membrane. The membrane conformations then can be described by the local separation $l_i$ of all pairs of apposing patches. Two apposing membrane patches can be pinned together by a single receptor-ligand bond. We assume here that the bonds are irreversible and rather rigid. At patch sites $i$ with receptor-ligand bonds, the local separation is constrained then to the length $l_i=l_o$ of the receptor-ligand bonds. 

The conformations of the two membranes are governed by the bending energy of the membranes, under the constraints that the local separation $l_i$ is equal to $l_o$ at all patch sites $i$ where receptor-ligand bonds are present. To simplify the notation, we use the rescaled separation field $z_i = (l_i/a)\sqrt{\kappa/k_BT}$, where $k_B$ is Boltzmann's constant, and $T$ is temperature.   Here, $\kappa = \kappa_1\kappa_2/(\kappa_1+\kappa_2)$ is the effective bending rigidity of the two membranes with rigidities $\kappa_1$ and $\kappa_2$.  The conformations of the rescaled separation field $z_i$ are governed by the effective bending energy \cite{Lipowsky88}
\begin{equation}
{\cal H}\{z\} = \sum_i {\textstyle \frac{1}{2}} \left(\Delta_d z_i\right)^2  \label{Hamiltonian}
\end{equation}
with $\Delta_{d}z_i =z_{i1}+z_{i2}+z_{i3}+z_{i4}-4 z_i$ where $z_{i1}$ to $z_{i4}$ are the membrane separations at the four nearest-neighbor patches of membrane patch $i$. The discretized Laplacian $\Delta_{d}z_i$ is proportional to the local mean curvature of the separation field. We focus here on the bending energy of the membranes since this energy is the dominant contribution to the elastic energy on the small length scales up to about 100 nm that are relevant here. Elastic contributions from a membrane tension $\sigma$ or from the cytoskeleton anchored to cell membranes are only important on larger length scales, i.e.~on length scales larger than the crossover length $\sqrt{\kappa/\sigma}$, which has a typical size of several hundred nanometers for rigidities $\kappa$ between 10 and 20 $k_BT$ \cite{Seifert95} and tensions $\sigma$ of a few $\mu$J/m \cite{Simson98}, and larger than the { average} separation of { neighboring} cytoskeletal anchor points, which is about 100 nm \cite{Alberts02}. 

\begin{figure}[t]
\begin{center}
\resizebox{0.8\columnwidth}{!}{\includegraphics{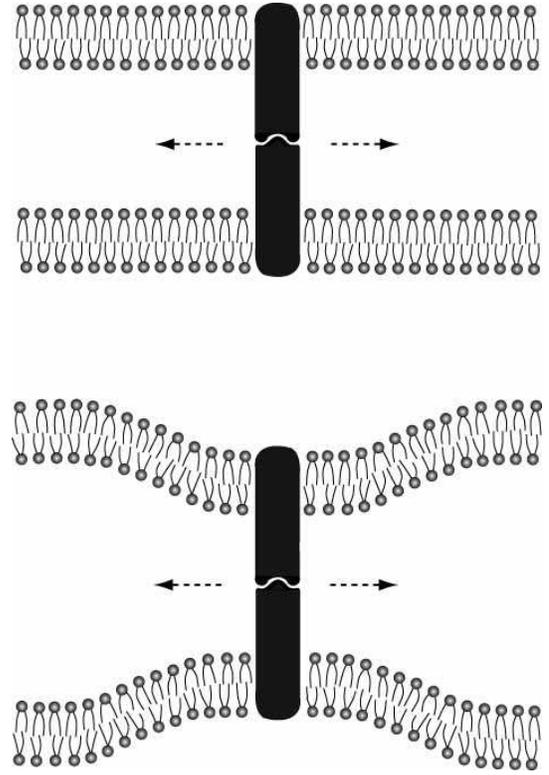}}
\caption{(Left) A receptor-ligand complex that interconnects two planar and parallel membranes. The lateral diffusion of the receptor-ligand complex, indicated by the arrows, does not involve any membrane bending. -- (Right) A receptor-ligand complex interconnecting two thermally fluctuating or `rough' membranes. The lateral diffusion is coupled to  the membrane fluctuations. }
\label{figure_cartoon}
\end{center} 
\end{figure}

The lateral diffusion of the receptor-ligand bonds is described as a hopping process on the square lattice of membrane patches. Let us first consider the hypothetical situation in which the membranes are planar and have a constant local separation $z_i = z_o$ where $z_o \equiv (l_o/a)\sqrt{\kappa/k_BT}$ is the rescaled length of the  receptor-ligand bonds. This situation is depicted in the left cartoon of Fig.~\ref{figure_cartoon}. The lateral diffusion of the receptor-ligand bonds is then independent of the membrane conformations. During each time step $t_d$ of the diffusive hopping process, each receptor-ligand bond can hop to one of the four nearest-neighbor patches, provided the patch is not occupied already by another receptor-ligand bond. At low concentrations of receptor-ligand bonds, diffusive encounters of two bonds can be neglected and the diffusion of a single receptor-ligand bond  is characterized by a sequence of `jump vectors' $\boldsymbol{a_i}$ with length $|\boldsymbol{a_i}|=a$. The orientation of each jump vector along one of the four `hopping directions' on the square lattice is independent of the orientations of the other jump vectors. After $N$ jumps, the lateral mean-square displacement $\boldsymbol{r}^2$ of $\boldsymbol{r} \equiv \sum_{i=1}^{N}  \boldsymbol{a_i}$ then simply is $\boldsymbol{r}^2= N a^2$, and the diffusion constant $D$, defined by $\boldsymbol{r}^2 \equiv 4 D t$ with $t=Nt_d$, is $D=a^2/4t_d$.

Lipid membranes, however, are not planar. They exhibit significant thermal shape fluctuations since their typical bending rigidities are between 10 and 20 $k_B T$ \cite{Seifert95}, i.e.~only one order of magnitude larger than the thermal energy $k_B T$. The fluctuations affect the lateral diffusion of the bound receptor-ligand complexes. The jumps of receptor-ligand bonds to neighboring membrane patches $j$ then depend on the local separation $z_j$ of these patches. We assume here that a jump, or hopping step, to a neighboring patch $j$ implies that the local separation $z_j$ of this patch is shifted to $z_o$, the rescaled length of the receptor-ligand bond. The hopping step then changes the bending energy (\ref{Hamiltonian}) of the membranes. We study the diffusion in Monte Carlo simulations with the standard Metropolis criterion \cite{Binder92}. According to this criterion, a hopping step is always accepted if the bending energy is reduced by the step. If the bending energy is increased by $\Delta E$, the step is accepted with probability $\exp[-\Delta E/k_B T]$. Hopping steps are always rejected if the neighboring membrane patch is already occupied by another receptor-ligand bond. The thermal fluctuations of the membranes are simulated via Monte Carlo steps that shift the local separation $z_i$ of patch $i$ to a `new' value $z_i + \rho$ where $\rho$ is a random number between -1 and 1. These local moves of the separation field are again accepted or rejected according to the Metropolis criterion, except for moves with $z_i+\rho<0$, which are always rejected since the membranes cannot penetrate each other.

An important aspect are the characteristic time scales for the lateral diffusion and for the membrane fluctuations on the smallest length scale $a\simeq 5$ nm of the model. Typical diffusion constants $D$ for membrane proteins are between 0.1 and 1 $\mu$m$^2/$s \cite{Kusumi05,Gambin06}, which correspond to characteristic jump times $t_d = a^2/4 D$ between 1 and a few microseconds. The characteristic relaxation time for membrane shape fluctuations on the length scale $a\simeq 5$ nm is around 0.1 microseconds \cite{Seifert94}, i.e.~one order of magnitude smaller than the characteristic diffusion time on this length scale. To capture this difference in the two time scales, we perform on average 10 Monte Carlo steps for the local separation $z_i$ on each lattice site between consecutive hopping steps of the receptor ligand-bonds. The local separation at the neighboring patches of a receptor-ligand bond `anneals' during these 10 Monte Carlo steps, i.e.~the separations are not, or only very weakly, time-correlated with the separations of these patches at the last hopping step of the receptor-ligand bond. 

The data presented here are from simulations with up to $10^7$ Monte Carlo steps per membrane patch for the separations $z_i$. We have simulated membranes with a size up to $120 \times 120$ patches and periodic boundary conditions. In our simulations, the lateral correlation length of the separation field is significantly smaller than the linear membrane size. Our results are therefore not affected by the finite membrane size. { The initial conformation of the Monte Carlo simulations is the planar membrane with separation $z_i=z_o$ and random distribution of receptor-ligand bonds. From this initial conformation, the membranes relax relatively fast into thermal equilibrium, for which the thermal averages reported below have been measured.} 

\begin{figure}[t]
\begin{center}
\onefigure{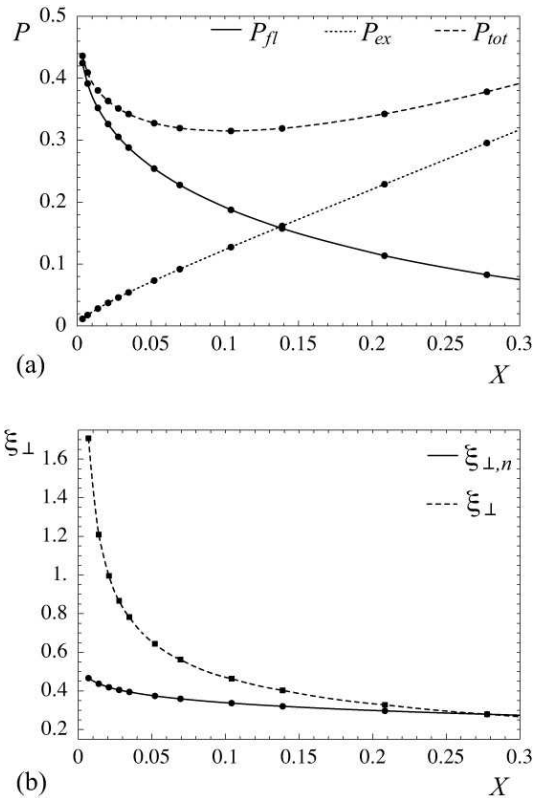}
\caption{(a) Rejection probabilities for lateral diffusion steps, or `hopping steps', as a function of the area fraction $X$ of receptor-ligand bonds in the limit of large rescaled bond length $z_o$.  The dotted line represents the probability $P_{ex}$ that a step is rejected because the membrane patch is already occupied by a receptor-ligand bond (`exclusion rejection'), the full line represents the probability $P_{fl}$ that a step is rejected because of an increase in the membranes' elastic energy (`fluctuation rejection'). The dashed line is the total rejection probability $P_{tot}$. The lines result from interpolation of the data points obtained in Monte Carlo simulations. The statistical errors for the the data points are smaller than the symbol sizes. 
-- (b) Overall membrane roughness $\xi_{\perp}$ (dashed line) and roughness $\xi_{\perp,n}$ at unoccupied neighbor sites of receptor-ligand bonds (full line) as a function of the bond area fraction $X$. The lines are obtained from interpolation of the Monte Carlo data points.}
\label{figure_rejection}
\end{center} 
\end{figure}
\begin{figure}[t]
\begin{center}
\onefigure{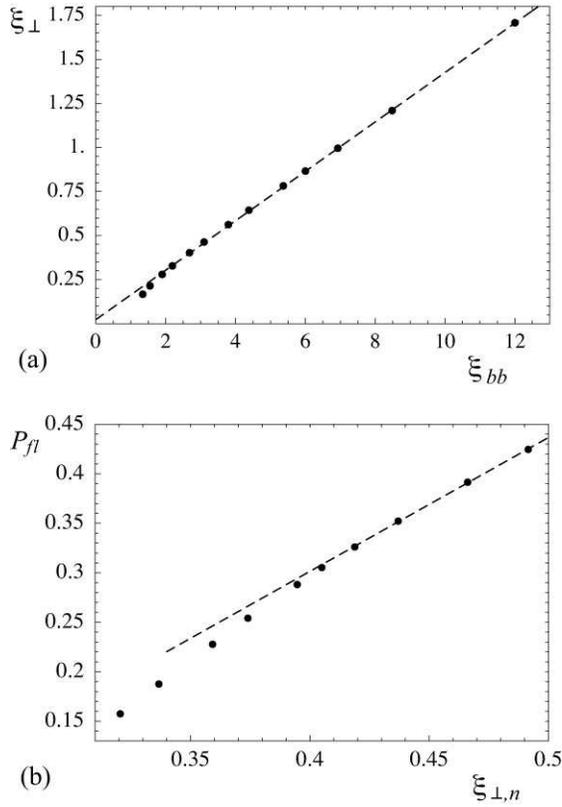}
\caption{(a) The overall membrane roughness $\xi_{\perp}$ for large rescaled bond lengths $z_o$ versus the mean bond separation $\xi_{bb}\equiv 1/\sqrt{X}$ in units of the linear size $a$ of membrane patches. Here, $X$ is the area fraction of receptor-ligand bonds. The line, given by $\xi_{\perp} = 0.02 + 0.14\, \xi_{bb}$, is obtained by a linear fit to the last four data points for large $\xi_{bb}$. -- (b) Fluctuation rejection probability $P_{fl}$ for lateral hopping steps versus the roughness $\xi_{\perp,n}$ at unoccupied nearest neighbor sites of bonds on the quadratic array of membrane patches. At large values of $\xi_{\perp,n}$, which correspond to small bond area fractions $X$, the rejection probability $P_{fl}$ is proportional to $\xi_{\perp,n}$. The dashed line $P_{fl} = -0.24 + 1.35 \,\xi_{\perp,n}$ is obtained by a linear fit to the four Monte Carlo data points for $\xi_{\perp,n}>0.41$.
}
\label{figure_roughness}
\end{center} 
\end{figure}
%

\section{Results}

Our model has only two parameters: (i) The area fraction $X$ of receptor-ligand bonds, i.e.~the fraction of membrane patches that contain bonds, and (ii) the rescaled length $z_o = (l_o/a)\sqrt{\kappa/k_BT}$ of the bonds, which depends on the actual bond length $l_o$, the effective bending rigidity $\kappa$ and the temperature $T$. { The characteristic length scale related to the area fraction is the mean bond separation $\xi_{bb}\equiv1/\sqrt{X}$. Other length scales, such as the average membrane separation $\langle z_i\rangle$  and the membrane roughness $\xi_{\perp} \equiv \sqrt{\langle z_i^2 \rangle - \langle z_i\rangle^2}$, depend on these two parameters (see below).} Here, $\langle \ldots\rangle$ denotes a { thermal} average over membrane conformations.  

We first consider the case where $z_o$ is significantly larger than the membrane roughness $\xi_{\perp}$. In the limiting case of large $z_o$, the fluctuating membranes don't `touch' each other, and the hard-wall repulsion at local separations $z_i=0$ does not play any role. Therefore, in this limit the results do not depend on $z_o$, i.e.~the area fraction $X$ of receptor-ligand bonds is the only parameter that affects the lateral diffusion of the bonds and the membrane roughness. In addition to the overall membrane roughness $\xi_\perp$ defined above, we will also consider the membrane roughness $\xi_{\perp,n}$ at unoccupied nearest-neighbor sites of bonds on the quadratic array of membrane patches. These unoccupied nearest-neighbor sites are the sites to which the bonds can jump during diffusion.

The effect of the membrane fluctuations on the lateral diffusion of the receptor-ligand bonds can be characterized by the rejection probabilities of hopping steps. There are two reasons why hopping steps can be rejected. First, a hopping step is rejected if the neighboring membrane patch is already occupied by a receptor-ligand bond. We call this `exclusion rejection', since the bonds exclude each other from the patches. The second reason is the `fluctuation rejection', i.e.~hopping steps are rejected with probability $\exp[-\Delta E/k_B T]$ if the step increases the bending energy of the fluctuating membranes by $\Delta E$. These two rejection probabilities and their sum, the total rejection probability, are shown in Fig.~\ref{figure_rejection}(a) as a function of the bond area fraction $X$  in the limit of large $z_o$. The exclusion rejection (dotted line) increases with the bond area fraction. The fluctuation rejection (full line), in contrast, decreases with the area fraction. The total rejection probability has a minimum at intermediate area fractions.

Concentrations of receptor-ligand bonds in cell-cell adhesion zones can be as large as several hundred bonds per $\mu$m$^2$, which corresponds to area fractions up to $0.01$ for membrane patch sizes $a\simeq 5$ nm. At these bond area fractions, the exclusion rejection is negligible. To understand the increase in the fluctuation rejection probability with decreasing bond fraction $X$, it is instructive to consider how the membrane roughness depends on $X$. The overall membrane roughness $\xi_{\perp}$, represented by the dashed line in Fig.~\ref{figure_rejection}(b) diverges with decreasing bond fraction $X$. We find that the roughness scales as $\xi_{\perp}\simeq 0.14 \,\xi_{bb}$ with $\xi_{bb}\equiv1/\sqrt{X}$, see Fig.~\ref{figure_roughness}(a). In contrast to the overall membrane roughness, the roughness $\xi_{\perp,n}$ at unoccupied nearest neighbor sites of receptor-ligand bonds (full line in Fig.~\ref{figure_rejection}(b)) reaches a finite value close to 0.5, or $0.5 a \sqrt{k_B T/\kappa}$ in physical units. At small area fractions $X$, the roughness $\xi_{\perp,n}$ is proportional to the fluctuation rejection probability for hopping steps, see Fig.~\ref{figure_roughness}(b). Therefore, the membrane rejection probability also reaches a finite value in the limit of small bond area fractions. This value is about 0.44, which implies that the numerical factor by which the diffusion constant of the receptor-ligand bonds is reduced due the thermal fluctuations is $1/(1-0.44) \simeq 1.8$ in the biologically relevant regime of small bond concentrations.

\begin{figure}[t]
\begin{center}
\onefigure{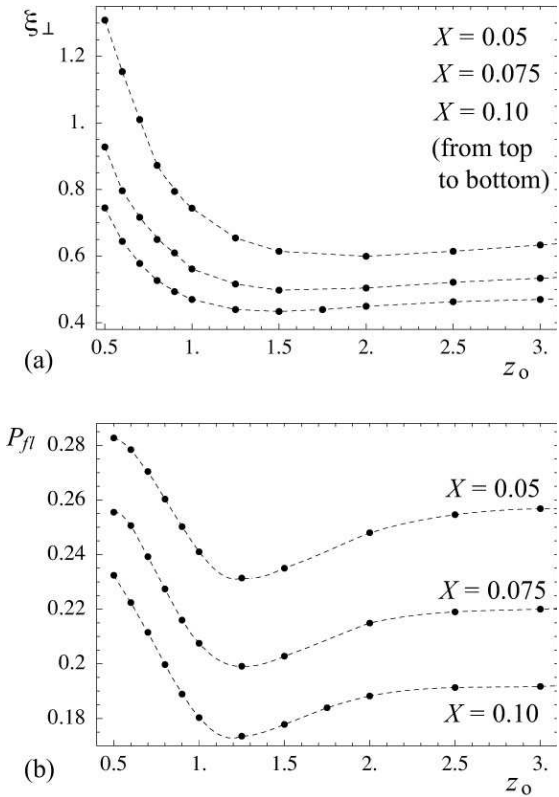}
\caption{(a) Overall membrane roughness $\xi_\perp$ as a function of the rescaled length $z_o$ of the receptor-ligand bonds at three different bond area fractions $X$, and: (b) Fluctuation rejection probability $P_{fl}$ versus $z_o$ at the same three bond area fractions $X$. For $z_o\lesssim 1.3$, the roughness $\xi_\perp$ and the fluctuation rejection probability $P_{fl}$ for hopping steps increase with decreasing $z_o$. }
\label{figure_bondlength}
\end{center} 
\end{figure}
\begin{figure}[t]
\begin{center}
\resizebox{0.87\columnwidth}{!}{\includegraphics{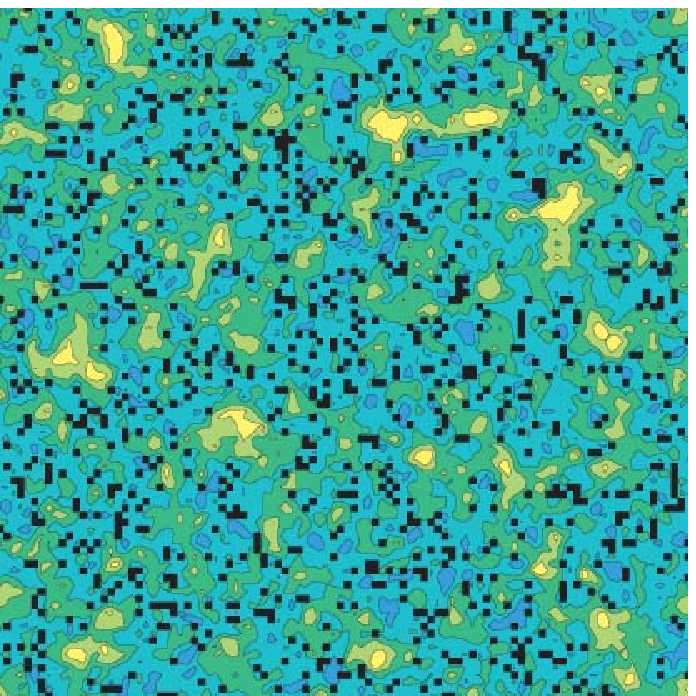}}

\vspace{0.5cm}

\resizebox{0.87\columnwidth}{!}{\includegraphics{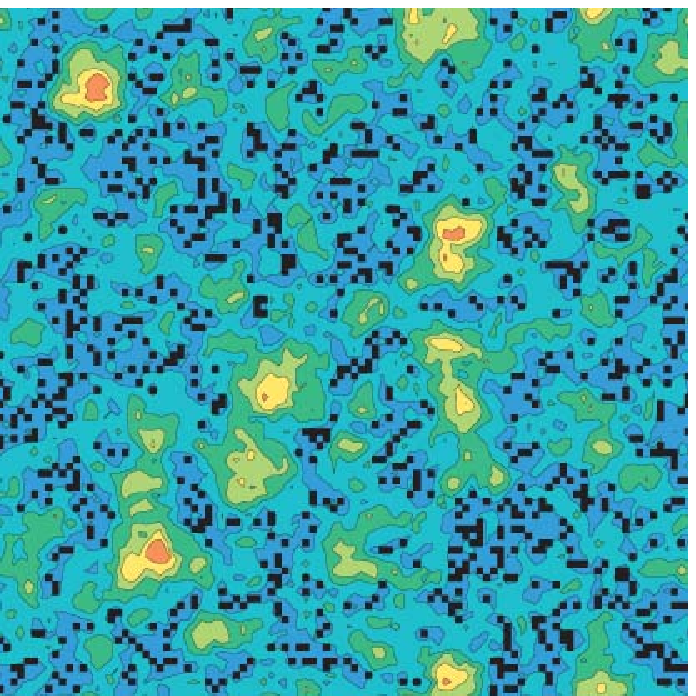}}
\caption{(Top) Monte Carlo configuration at the concentration $X=0.1$ of receptor-ligand bonds in the limit of large rescaled bond length $z_o$. Bonds are shown in black, and the local membrane separation is indicated by colors ranging from blue (small separations) over green to yellow (large separations). The membrane consists of $100\times 100$ patches of size $a^2$. -- (Bottom) Monte Carlo configuration at the same concentration $X=0.1$ and the relatively small rescaled bond length $z_o=0.5$. The membrane separation $z$ varies between 0 (blue color) and 4.85 (red color).  The clustering of receptor-ligand bonds is caused by the fluctuation-induced repulsion of the membranes.
  }
\label{figure_snapshots}
\end{center} 
\end{figure}

The rescaled bond length $z_o$ can affect the diffusion if it is close to or smaller than the roughness $\xi_{\perp}$. Typical receptor-ligand bonds that interconnect cell membranes have a linear extension $l_o$ between 15 and 40 nm. For bending rigidities $\kappa$ from 10 to 20 $k_BT$, this corresponds to rescaled bonds lengths $z_o=l_o/a\sqrt{\kappa/k_B T}$ between 10 and 40 at room temperature. For typical bond area concentrations $X/a^2$ in cell adhesion zones between 10 and several hundred molecules per $\mu$m$^2$, the roughness $\xi_{\perp}\simeq 0.14/\sqrt{X}$ is smaller than these rescaled bond lengths. The length of the bonds then does not affect the diffusion behavior. However, biomimetic, or `artificial', receptor-ligand systems can be significantly smaller than the biological bonds. For example, membrane bonds between negatively charged lipids mediated by multivalent ions such as Cr$^{3+}$ can impose local membrane separations smaller than 1 nm \cite{Franke06}. Fig.~\ref{figure_bondlength} shows how the overall membrane roughness and the membrane rejection probability $P_{fl}$ for hopping steps depend on the length of relatively short bonds. As a function of the bond length, the roughness has a minimum at intermediate lengths, and increases for small lengths. This increase is a consequence of the fluctuation-induced repulsion of the membranes. For small bond lengths $z_o$, the average membrane separation is then significantly larger than $z_o$, which causes an increase in the roughness.  This is reminiscent of the increase in roughness with decreasing bond length in the case of membranes adhering via parallel stripes of receptor-ligand bonds \cite{Weikl00b}. The fluctuation-induced repulsion of the membranes induce also attractive interactions between the bonds { (see Fig.~\ref{figure_snapshots})}. These entropic attractions can lead to lateral phase separation for large bonds that occupy several membrane patches \cite{Weikl00} or for bonds that locally change the rigidity of membrane patches in which they are anchored \cite{Weikl01}, but not for the small bonds considered here \cite{Weikl00}.  { Aggregation of receptor-ligand bonds in the adhesion zones of cells \cite{Grakoui99,Delanoe04} or vesicles \cite{Boulbitch01} has also been observed experimentally. In these situations, the aggregation is presumably driven by a length mismatch (i) between bonds and repulsive molecules \cite{Boulbitch01} or (ii) between} { different types of receptor-ligands bonds \cite{Grakoui99}, which leads to a barrier in the effective adhesion potential of the membranes \cite{Weikl01}.}

\section{Discussion and conclusions}

We have focused here on the effect of membrane shape fluctuations on the diffusion of receptor-ligand bonds between two membranes. We have found that the fluctuations reduce the bond diffusion constant by a factor close to 2 at small bond area fractions, compared to planar membranes. Membrane fluctuations can also affect observed diffusion constants via the ratio of the actual and the projected membrane area in the experimental focal plane \cite{Reister05,Gov06}, by coupling to the spontaneous curvature of membrane proteins \cite{Reister05}, or by curvature-induced changes of the membrane thickness \cite{Gov06}. As `pinning centers' between fluctuating membranes, the bonds considered here can also impede the lateral diffusion of other large membrane proteins \cite{Lin04}.

We have assumed that the receptor-ligand bonds in our model are rather rigid and have a constant bond length $l_o$. This assumption is justified as long as variations in the bond length are significantly smaller than the roughness $\xi_{\perp,n}$ at bond neighbor patches. At small bond area fractions, $\xi_{\perp,n}$ is close to 0.5, which corresponds to variations $\xi_{\perp,n} a \sqrt{k_BT/\kappa}$ in the membrane separation up to 1 nm. Another assumption of our model is that Monte Carlo diffusion steps of the receptor-ligand bonds imply a shift of the local membrane separation and an associated change of the membrane bending energy. { 
We have previously shown that the Monte Carlo dynamics for the membrane separation field $z$ corresponds to a Langevin dynamics in the limit of small Monte Carlo step widths $\delta z$ \cite{Rozycki06b}. For the diffusion steps, we have chosen here for simplicity a finite Monte Carlo step width that is identical with the lattice spacing $a$ of the discrete membrane, which prevents the mapping to a continuous Langevin dynamics. We have also neglected hydrodynamic interactions, which may lead to a further reduction of the diffusion constant. 
Theoretical descriptions of diffusion in membranes modeled as elastic sheets require assumptions and approximations that, in principle, can be elucidated in simulations with molecular membrane models \cite{Goetz99,Guigas06}.} Our theoretical results are also accessible to experimental studies of model systems with irreversibly bound receptor-ligand complexes such as the Biotin-Avidin complex.

\begin{acknowledgements}
H.\ K.\ and G.\ J.\ S.\  acknowledge support  by the Austrian Science Fund (FWF) and the Wilhelm-Macke Travel scholarship.
\end{acknowledgements}

\end{document}